\documentclass[aps,showpacs,showkeys,twocolumn,footnoteinbib]{revtex4}
\usepackage{mathrsfs}

\usepackage{epsfig}
\usepackage{amssymb}
\usepackage{bm}
\usepackage{hyperref}
\usepackage{color}
\usepackage{amsmath}

\begin{document}

\title{Chiral Symmetry Breaking in the Dynamical Soft-Wall Model}
\author{Qi Wang$^{1}$\footnote{Email:anomalyfree@gmail.com}, An Min Wang$^{2}$\footnote{Email:anmwang@ustc.edu.cn}\\}
\vspace{0.5cm} \affiliation{
\bigskip
$^1$Department of Modern Physics, University of Science and
Technology of China, Hefei 230026, PR China\\
$^2$Department of Modern Physics, University of Science and
Technology of China, Hefei 230026, PR China}


\begin{abstract}
In this paper a model incorporating chiral symmetry breaking and
dynamical soft-wall AdS/QCD is established. The AdS/QCD background
is introduced dynamically as suggested by Wayne de Paula etc al and
chiral symmetry breaking is discussed by using a bulk scalar field
including a cubic term. The mass spectrum of scalar, vector and
axial vector mesons are obtained and a comparison with experimental
data is presented.

\end{abstract}

\maketitle

\section{Introduction}
\label{secIntro}

Quantum Chromodynamics (QCD) is an non-abelian gauge theory of the
strong interactions. It has a coupling constant that highly depends
on the energy scale, and the perturbation theory doesn't work at low
energies. The Anti-de Sitter/conformal field theory (AdS/CFT)
correspondence~\cite{Maldacena:1997re} provides a novel way to
relate QCD with a 5-D gravitational theory and calculate many
observables more efficiently. This leads to two types of AdS/QCD
models: the top-down models that are constructed by branes in string
theory and the bottom-up models by the phenomenological aspects. In
the top-down model, mesons are identified as open strings with both
ends on flavor branes~\cite{Karch:2002sh}. The most successful
top-down model in reproducing various facts of low energy QCD
dynamics is D4-D8 system of Sakai and Sugimoto
~\cite{Sakai:2004cn}~\cite{Sakai:2005yt}.

The bottom-up approach assumes that QCD has suitable 5D dual and
construct the dual model from a small number of operators that are
influential. The first bottom-up model is the so-called hard-wall
model~\cite{EKSS}~\cite{DP}; it simply place an infrared cutoff on
the fifth dimension to break the chiral symmetry. But in hard-wall
model the spectrum of mass square $m_n^2$ grow as $n^2$, which is
contradict with the experimental data~\cite{Shifman-Regge}. The
analysis of experimental data indicates a Regge behavior of the
highly resonance $m_n^2\sim n$. To compensate this difference
between QCD and AdS/QCD, soft-wall model was
proposed~\cite{Karch:2006pv}. In this model the spacetime smoothly
cap off instead of the hard-wall infrared cutoff by introducing a
background dilaton field $\Phi$: $$S_{5}=-\int d^{5}x
\sqrt{-g}\,e^{-\Phi(z)}\mathscr{L},$$where the background field is
parametrized to reproduce Regge-like mass spectrum. In spite of
success in reproducing Regge-like mass spectrum, chiral symmetry
breaking in this model is not QCD-like. Some further researches try
to incorporate chiral symmetry breaking and confinement in this
model by using high order terms in the potential for scalar
field~\cite{Gherghetta:2009ac}~\cite{zhangpeng:2010prd}.

But the dilaton field and scalar field in the soft-wall model are
imposed by hand to and cannot be derived from any equation of
motion. Some works on the dilaton-gravity coupled model show that a
linear confining background is possible as a solution of the
dilation-gravity coupled equations and Regge trajectories of meson
spectrum is obtained~\cite{dePaulaPRD09}~\cite{dynamicalmetric}.

In this paper, we try to incorporate soft-wall AdS/QCD and dynamical
dilaton-gravity model. We introduce the meson sector action with a
cubic term of the bulk field in the bulk scalar potential under the
dynamical metric background. We derive a nonlinear differential
equation related the vacuum expectation value(VEV) of bulk field,
wrap factor and dilaton field. An analysis of the asymptotic
behavior of the VEV is presented. We obtain that the VEV is a
constant in the IR limit that suggests chiral symmetry restoration
exists. The coupling constant can be determined by other parameters
in this model.

This paper proceeds as follow: A brief review of dynamical
gravity-dilaton background is presented in Sec. \ref{secMetric}. In
Sec. \ref{model} we introduce the bulk field and meson sector action
under dynamical metric background and obtain the differential
equation of the bulk field VEV. An analysis of asymptotic behavior
is presented and a parametrized solution is given. Using the
parametrized solution we calculate the scalar, vector and axial
vector mass spectrums in Sec. \ref{massspectrum}. A conclusion is
given in Sec. \ref{conclusion}.


\section{Background Equations}
\label{secMetric}

We start from the Einstein-Hilbert action of five-dimensional
gravity coupled to a dilaton $\Phi $ as proposed by de Paula,
Frederico, Forkel and Beyer \cite{dePaulaPRD09}:

\begin{equation}
S=\frac{1}{2\kappa ^{2}}\int d^{5}x\sqrt{-g}\left( -%
\emph{R}+\frac{1}{2}g^{MN}\partial _{M}\Phi \partial _{N}\Phi
-V(\Phi )\right)  \label{actiongd}
\end{equation}
where $\kappa $ is the five-dimensional Newton constant and $V$ is a
still general potential for the scalar field. Then we restrict the
metric to the form
\begin{equation}
g_{MN}=e^{-2A(z)}\eta_{MN} \label{metric}
\end{equation}
where $\eta_{MN}$ is the Minkowski metric.  We write the warp factor
as
\begin{equation}
A(z)=\ln z+C\left( z\right)  \label{wrapfactor}
\end{equation}

Variation of the action (\ref{actiongd}) leads to the
Einstein-dilaton equations for the background fields $A$ and $\Phi$:
\begin{eqnarray}
6A^{\prime }{}^{2}-\frac{1}{2}\Phi ^{\prime 2}+e^{-2A}V(\Phi ) &=&0
\label{einsteinzz} \\
3A^{\prime \prime }-3A^{\prime }{}^{2}-\frac{1}{2}\Phi ^{\prime
2}-e^{-2A}V(\Phi ) &=&0  \label{einstein00} \\
\Phi ^{\prime \prime }-3A^{\prime }\Phi ^{\prime
}-e^{-2A}\frac{dV}{d\Phi } &=&0  \label{dilatonequation}
\end{eqnarray}
Here the prime denotes the derivative with respect to $z$.

By adding the two Einstein equation ones can get the dilaton field
directly:
\begin{equation} \Phi ^{\prime }=\sqrt{3}\sqrt{A^{\prime
}{}^{2}+A^{\prime \prime }}~ \label{constrain}
\end{equation}

and by substituting (\ref{constrain}) in (\ref{einsteinzz}) or
(\ref{einstein00}):
\begin{equation}
V(\Phi \left( z\right) )=\frac{3e^{2A\left( z\right) }}{2}\left[
A^{\prime \prime }\left( z\right) -3A^{\prime }{}^{2}\left( z\right)
\right] \label{vz}
\end{equation}

If  $A(z)=\log(z)+z^\lambda$ is set for simplicity, the asymptotic
behavior of dilation field $\Phi$ in UV and IR limits can be
obtained

\begin{equation}
\Phi(z)=\left\{
         \begin{array}{ll}
\underrightarrow{z\rightarrow0} & 2\sqrt{3(1+\lambda^{-1})}z^{\frac{\lambda}{2}}\\
\underrightarrow{z\rightarrow\infty} & \sqrt{3}z^\lambda
         \end{array}
         \right.
\end{equation}


\section{The Model}
\label{model}

In this section, we will establish the soft-wall AdS/QCD model first
introduced in~\cite{Karch:2006pv} under the dynamical background
mentioned in Sec. \ref{secMetric}. The background geometry is chosen
to be 5D AdS space with the metric
\begin{equation}
ds^{2}=g_{MN}dx^{M}dx^{N}=e^{-2A}\left(
\eta_{\mu\nu}dx^{\mu}dx^{\nu}+dz^{2}\right)~ \label{background}
\end{equation}
where $A$ is the warp factor, and $\eta_{\mu\nu}$ is Minkowski
metric. Unlike~\cite{Karch:2006pv} we introduce a background dilaton
field $\Phi$ that is coupled with $A$ and $\Phi$ can be determined
in terms of $A$ as shown in Sec. \ref{secMetric}.

In order to describe chiral symmetry breaking, a bifundamental
scalar field $X$ is introduced and we add an cubic term potential
instead of the quartic term as shown in~\cite{Gherghetta:2009ac}
\begin{equation}
S_{5}=-\int d^{5}x \sqrt{-g}\,e^{-\Phi(z)}{\rm
Tr}[|DX|^{2}+m_{X}^{2} |X|^{2} \nonumber
\end{equation}
\begin{equation}
-\lambda|X|^{3}+\frac{1}{4g_{5}^{2}}(F_{L}^{2}+F_{R}^{2})]
\label{mesonaction}
\end{equation}
where $m_{X}=-3$, $\lambda$ is a constant and
$g_5^2=12\pi^2/N_c=4\pi^2$. The fields $F_{L,R}$ are defined by
\begin{equation}
F_{L,R}^{MN} = \partial^{M} A_{L,R}^{N} - \partial^{N} A_{L,R}^{M} -
i [A_{L,R}^{M},A_{L,R}^{N}], \nonumber
\end{equation}
here $A_{L,R}^{MN} = A_{L,R}^{MN} t^{a}$, $Tr[t^{a} t^{b}] =
\frac{1}{2} \delta^{ab}$, and the covariant derivative is $D^M
X=\partial^M X-i A_L^MX+iXA_R^M$.

\subsection{Bulk scalar VEV solution and dilaton field}
The scalar field $X$ is assumed to have a z-dependent VEV
\begin{equation}
<X>=\frac{v(z)}{2}\left(\begin{array}{cc}1 & 0\\
                                         0 & 1
                        \end{array}\right)   \label{xvev}
\end{equation}

We can obtain a nonlinear equation related $A$, $\Phi$ and $v(z)$
from the action (\ref{mesonaction})
\begin{equation}
\partial_z( e^{-3A} e^{-\Phi} \partial_z v(z))+ e^{-5A} e^{-\Phi} (3 v(z)+\frac{3}{4}\lambda v^2(z))=0  \label{VEVequation}
\end{equation}

If $A$ and $\Phi$ are given, this equation can be simplified to a
second order differential equation of $v(z)$
\begin{equation}
v''(z)-(\Phi'+3A')v'(z)+e^{-2A}(3 v(z)+\frac{3}{4}\lambda v^2(z))=0
\label{VEVequation1}
\end{equation}
where $ \Phi ^{\prime }=\sqrt{3}\sqrt{A^{\prime }{}^{2}+A^{\prime
\prime }}~ \label{constrain} $.

Firstly, we consider the limit $z\rightarrow \infty$. If wrap factor
$C(z)$ has a asymptotic behavior of $z^2$ as $z\rightarrow \infty$,
we can determine that $v(z)\rightarrow$ constant. This means that
chiral symmetry restoration in the mass spectrum. But there are many
controversies on whether such a restoration really
exists~\cite{Cohen:2005am}~\cite{Wagenbrunn:2006cs}.

Secondly, we analyze the asymptotic behavior of $v(z)$ as
$z\rightarrow 0$. As shown
in~\cite{Witten:1998qj}~\cite{Klebanov:1999tb}, the VEV as
$z\rightarrow 0$ is required to be
\begin{equation}
v(z)=m_q\zeta z+\frac{\sigma}{\zeta}z^3
\end{equation}
where $m_q$ is the quark mass, $\sigma$ is the chiral condensate and
$\zeta=\frac{\sqrt{3}}{2\pi}$.

We assume that $C(z)$ takes the asymptotic form as $z\rightarrow 0$
\begin{equation}
C(z)=\alpha z^2+\beta z^3
\end{equation}
Then we can get the asymptotic form of $\Phi'$ in the IR limit.
Considering the coefficients of $\frac{1}{z}$ and constant terms at
the left side of Equation (\ref{VEVequation1}) should be eliminated
as $z\rightarrow 0$, we can conclude that the cubic term of bulk
field is necessary and a relation between $\alpha$, $\beta$ and
$\lambda$ can be obtained
\begin{equation}
\lambda m_q\zeta=4\sqrt{2\alpha}.  \label{lambdarelation}
\end{equation}

\subsection{A parametrized solution and parameter setting}
The equation (\ref{VEVequation1}) is a nonlinear differential
equation and it is difficult to solve  the VEV $v(z)$, so we choose
to select a parametrized form of $v(z)$ that satisfies the
asymptotic constraint instead of solving the equation directly.

We set the wrap factor as suggested in~\cite{dePaulaPRD09}
\begin{equation}
A(z)=\log z+\frac{(\xi\Lambda_{QCD} z)^2}{1+\exp(1-\xi\Lambda_{QCD}
z)} \label{wrapfactor}
\end{equation}
where $\xi$ is a scale factor and $\Lambda_{QCD}=0.3\textrm{GeV}$ is
the QCD scale.

Also we assume the VEV $v(z)$ asymptotically behaves as discussed in
previous section
\begin{equation}
v(z)=\left\{
          \begin{array}{ll}
\underrightarrow{z\rightarrow\infty} & \gamma\\
\underrightarrow{z\rightarrow0} & m_q\zeta z+\frac{\sigma}{\zeta}z^3
         \end{array}
         \right.  \label{VEVasymptotic}
\end{equation}

A simple parametrized form for $v(z)$ that satisfies asymptotic
constraint (\ref{VEVasymptotic}) can be chosen as
\begin{equation}
v(z)=\frac{z(A+Bz^2)}{\sqrt{1+(Cz)^6}}
\end{equation}
The relations between $m_q$, $\sigma$, $\gamma$ and $A$, $B$, $C$
are:
\begin{equation}
A=m_q\zeta, B=\frac{\sigma}{\zeta}, \frac{B}{C^3}=\gamma
\end{equation}

Using the data of meson mass data, pion mass and pion decay constant
we can get a good fitting of $A$, $B$, $C$, $\lambda$ as follows:

\begin{equation}
A=1.63\textrm{MeV}, B=(157\textrm{MeV})^3, C=10,
\lambda=\frac{4\sqrt{2}\xi \Lambda_{QCD}}{m_q\zeta\sqrt{1+e}}
\end{equation}

In next section we will use this parametrized solution to calculate
scalar, vector and axial-vector meson mass spectra and compare them
with experimental data.


\section{Meson Mass Spectrum}
\label{massspectrum}

Starting from the action (\ref{mesonaction}), we can derive the
Schr$\ddot{o}$dinger-like equations that describe the scalar, vector
and axial vector mesons:
\begin{equation}
-\partial_z^2S_n(z)+(\frac{1}{4}\omega'^2-\frac{1}{2}\omega''-e^{-2A}(3+\frac{3}{2}\lambda
v(z)))S_n(z)=m^2_{S_n}S_n(z)  \nonumber
\end{equation}

\begin{equation}
-\partial_z^2V_n(z)+(\frac{1}{4}\omega'^2-\frac{1}{2}\omega'')V_n(z)=m^2_{V_n}V_n
\nonumber
\end{equation}
\begin{equation}
-\partial_z^2A_n(z)+(\frac{1}{4}\omega'^2-\frac{1}{2}\omega''+g_5^2e^{-2A}v^2(z))A_n(z)=m^2_{A_n}A_n(z)
\label{mesonquation}
\end{equation}
where the prime denotes the derivative with respect to $z$ and
$\omega=\Phi(z)+3A(z)$ for scalar mesons, $\omega=\Phi(z)+A(z)$ for
vector and axial vector mesons.

The meson mass spectrum is obtained through calculating the
eigenvalues of these equations. We will calculate the eigenvalues of
these equations numerically using the parametrized solution
introduced in Sec. \ref{model}.

Since the potential is complicated, we must solve the eigenvalue
equation numerically. As suggest in~\cite{dePaulaPRD09}, we set
$\xi=0.58$ for scalars and the numerical results are shown in Table
I.

\begin{table}[ht]
\begin{center}
\caption{Scalar mesons spectra in MeV.}
\begin{tabular}{|c c|c|c|}
  \hline
  n & & $f_0 (Exp)$ & $f_0 (Model)$ \\
  \hline
  0 & & 550 & 897 \\
  \hline
  1 & & 980 & 1135 \\
  \hline
  2 & & 1350 & 1348 \\
  \hline
  3 & & 1505 & 1540 \\
  \hline
  4 & & 1724 & 1717 \\
  \hline
  5 & & 1992 & 1881 \\
  \hline
  6 & & 2103 & 2034 \\
  \hline
  7 & & 2134 & 2178 \\
  \hline
\end{tabular}
\end{center}
\end{table}

For vectors, the value of $\xi$ is chosen to be 0.88 and the
numerical results are listed in Table II.

\begin{table}[ht]
\begin{center}
\caption{Vector mesons spectra in MeV.}
\begin{tabular}{|c c|c|c|}
  \hline
  n & & $\rho (Exp)$ & $\rho (Model)$ \\
  \hline
  0 & & 775.5 & 988.9 \\
  \hline
  1 & & 1465 & 1348 \\
  \hline
  2 & & 1720 & 1650 \\
  \hline
  3 & & 1909 & 1913 \\
  \hline
  4 & & 2149 & 2147 \\
  \hline
  5 & & 2265 & 2359 \\
  \hline
\end{tabular}
\end{center}
\end{table}

Sine $\Delta m^2=m^2_{A^2_n}-m^2_{V^2_n}=g^2_5e^{-2A}v^2(z)$ tends
to zero as $z\rightarrow 0$, it means the mass of vector and
axial-vector mesons are equal at high values.


\section{Conclusion}
\label{conclusion} In this paper we incorporated chiral symmetry
breaking into the soft-wall AdS/QCD model with a dynamical
background. Then a discussion about the asymptotic behavior of the
VEV of the bulk field was presented and a chiral symmetry
restoration was suggested in the IR limit. We also introduced a
parametrized solution of the VEV of the bulk field and fitted the
parameters by using the meson mass spectra and pion mass and its
decay constant. The numerical solution of meson mass spectra and
comparison with experimental data was also considered. The agreement
between the theoretical calculation and experimental data is good.

There are some issues worthy of further consideration. The
differential equation of the VEV $v(z)$ needs further study,
numerically and analytically, to obtain a more precise form of
$v(z)$. The mass spectra of nucleons and some more complicated
properties of mesons and nucleons in this model will be our
following work.

\section{Acknowledgement}

This work has been supported by the National Natural Science
Foundation of China under Grant No. 10975125.

\end{document}